\documentclass[11pt]{article}
\usepackage{moriond,epsfig}
\begin{document}
\vspace*{1cm}
\title{Higgs Physics at a Future {\boldmath ${\mathrm e^+e^-}$} Linear Collider}
\author{Markus Schumacher}
\address{Deutsches Elektronensynchrotron, DESY, Notkestr.85, 22603 Hamburg, Germany}
%%%%%%%%%%%%%%%%%%%%%%%%%%%%%%%%%%%%%%%%%%%%%%%%%%%%%%%%%%%%%%%%%%%%%%%%%%%%%%%%%%%%%%%%%%%%%%
\maketitle\abstracts{
This letter reviews the potential of a high luminosity ${\mathrm e^+e^-}$ 
linear collider (LC) in the precision study of the Higgs boson profile.
The complementarity with the Large Hadron Collider 
(LHC) Higgs physics program is briefly discussed.}
%%%%%%%%%%%%%%%%%%%%%%%%%%%%%%%%%%%%%%%%%%%%%%%%%%%%%%%%%%%%%%%%%%%%%%%%%%%%%%%%%%%%%%%%%%%%%%
\section{Introduction: The Quest for the Higgs}
Unraveling the mechanism responsible for electroweak symmetry breaking 
and the generation of particle masses is one of the great scientific quests of high 
energy physics today. The Standard Model (SM) and its Supersymmetric (SUSY) extensions
address this question by the Higgs mechanism~\cite{higgs}. 
The first manifestation of the Higgs mechanism 
is represented by the existence of at least one Higgs boson. This motivates the large 
experimental effort for the Higgs boson search in the past and future decades. The findings
of the finished LEP2 program are a preliminary lower limit on the SM Higgs boson mass 
of 114.1\,GeV and 
the observation of a slight excess around 115\,GeV, compatible with the 
background-only-hypothesis at a level of 3.4\%~\cite{lep}. The hunt will continue at the Tevatron 
${\mathrm p\bar{p}}$-collider at $\sqrt{s}$ $\approx$ 2 TeV, covering the mass range  
with 3$\sigma$ 
evidence up to 180 GeV with 30\,fb$^{-1}$ per experiment~\cite{tevatron}. 
In ${\mathrm pp}$-collisions at $\sqrt{s}$ = 14\,TeV  at the LHC the entire SM mass range will be covered 
with at least 5$\sigma$ evidence with 30\,fb$^{-1}$ per experiment~\cite{lhc}. 
For the MSSM Higgs sector, at least one Higgs boson can be observed for the 
entire parameter space with an integrated luminosity of 300\,fb$^{-1}$~\cite{lhc}. 
However, the discovery at these hadron machines relies on the identification of specific Higgs 
decay modes.
At the LC the Higgs boson can be observed in the Higgs-strahlung process 
${\mathrm e^+e^- \rightarrow ZH}$
with ${\mathrm Z \rightarrow \ell^+ \ell^-}$, independently of its decay mode by a distinctive peak in the 
di-lepton recoil mass distribution. A data set of 500\,fb$^{-1}$ at $\sqrt{s}$ = 350\,GeV,
corresponding to 2 to 3 years of running, provides a sample of 4600-2300 Higgs particles for 
$M_{\mathrm H}$ between 120 and 200 GeV in this channel~\cite{tdr}. 

After the discovery the full validation of the Higgs mechanism requires an accurate
determination of the Higgs bosons production and decay properties. The potential of the LC
in the precision study of the Higgs boson profile is reviewed in Section2~\footnote{The results
discussed here are mainly based on the ECFA/DESY study as summarised in \cite{tdr}. In the meantime
similar studies have been performed and documented in other regional 
Higgs working groups for SNOWMASS 2001~\cite{snowmass}.}. 
The prospects for the investigation of the MSSM Higgs sector are discussed in Section 3
and finally the complementarity with the LHC is outlined in Section 4.

%%%%%%%%%%%%%%%%%%%%%%%%%%%%%%%%%%%%%%%%%%%%%%%%%%%%%%%%%%%%%%%%%%%%%%%%%%%%%%%%%%%%%%%%%%%%%%
\newpage
\section{Higgs Boson Profile}
\subsection{Higgs Mass}
\begin{figure}[htb]
\center{
\epsfig{figure=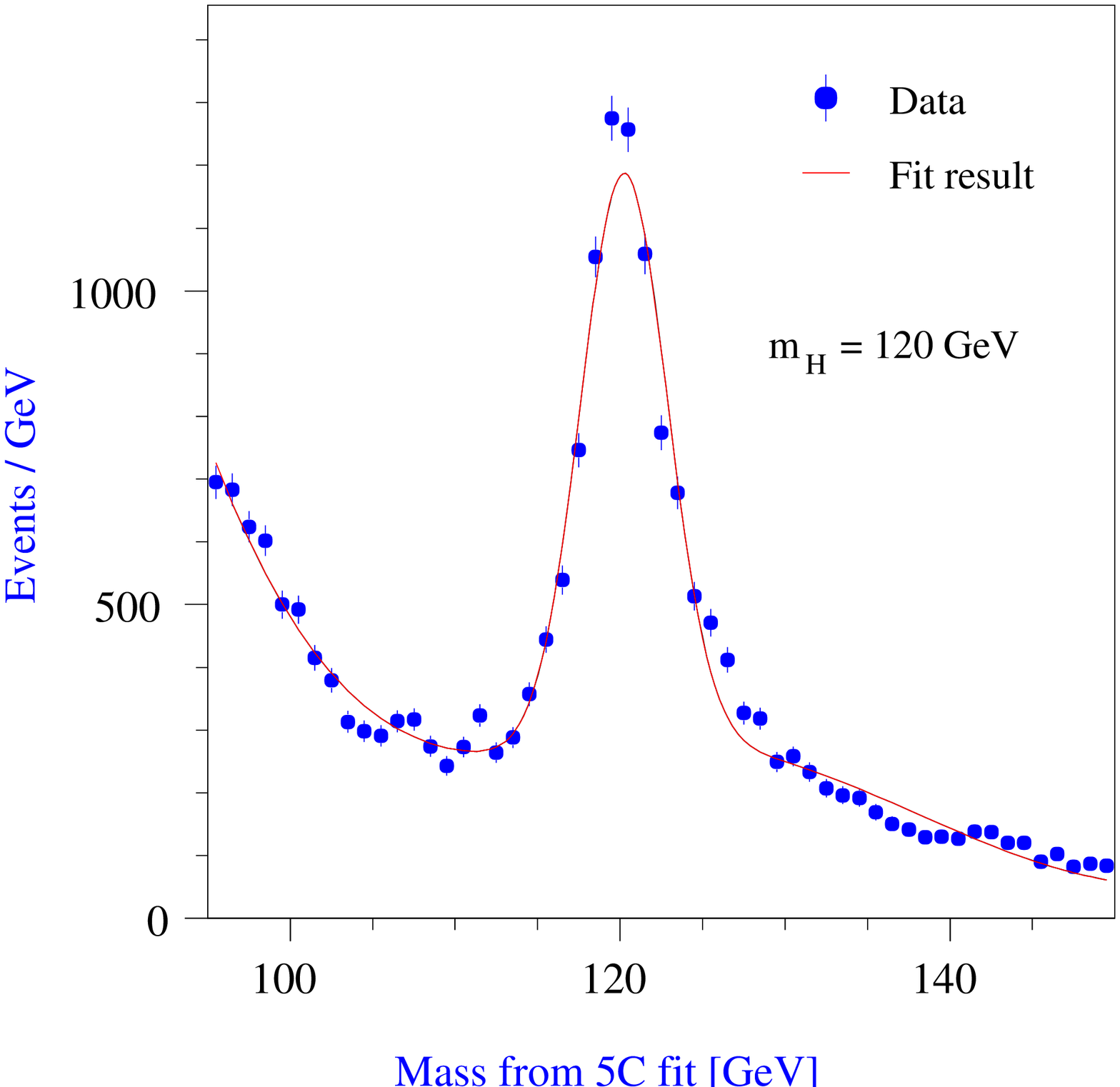,height=5.5cm,width=5.5cm}
\hspace{2cm}
\epsfig{figure=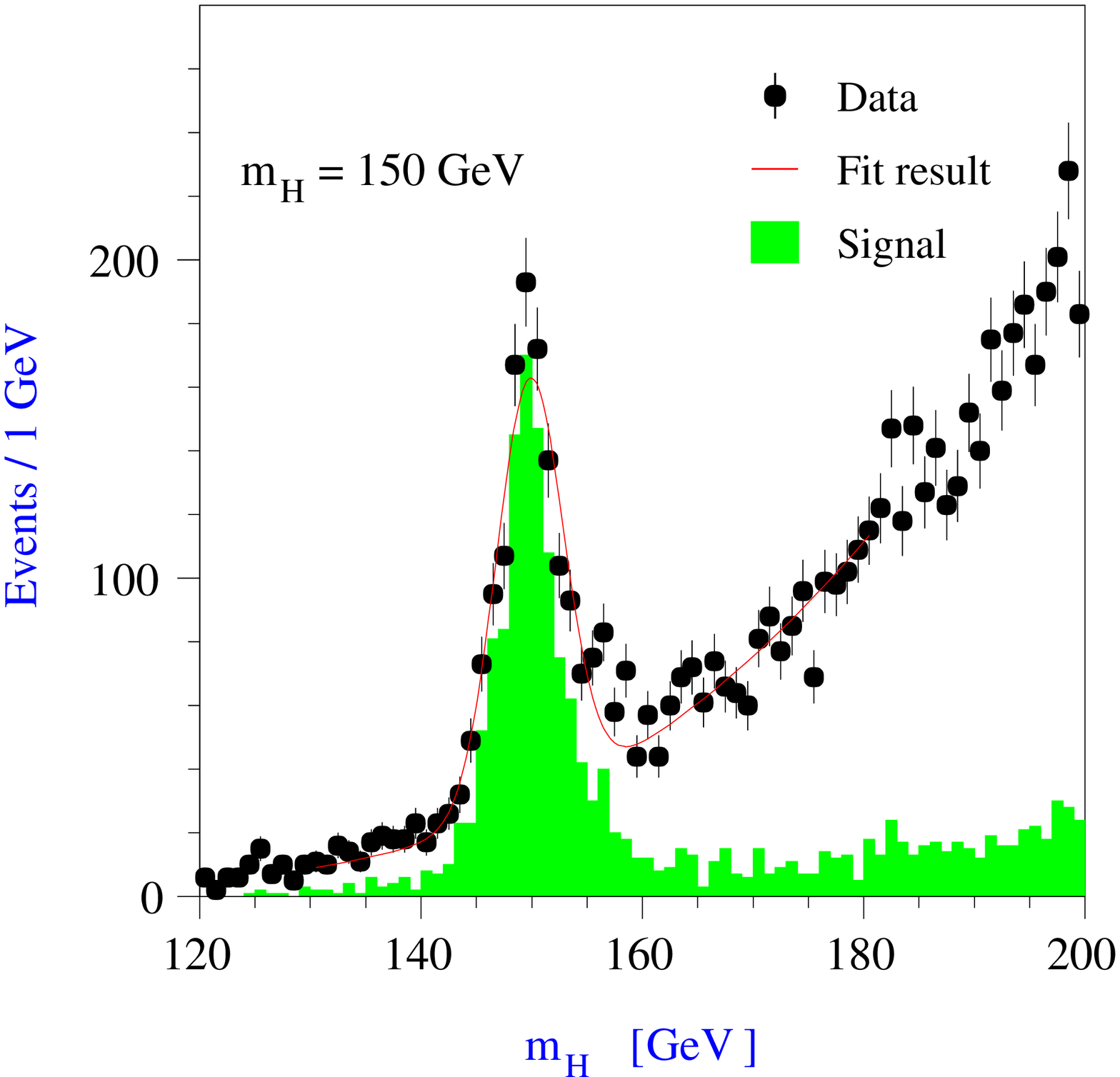,height=5.5cm,width=5.5cm}}
\caption{The Higgs boson mass reconstructed in the ${\mathrm H \rightarrow b\bar{b}}$, 
${\mathrm Z \rightarrow q\bar{q}}$ channel for $M_{\mathrm H}$ = 120\,GeV (left) and in the ${\mathrm H \rightarrow WW^*}$, ${\mathrm Z \rightarrow q\bar{q}}$ 
channel for $M_{\mathrm H}$ = 150\,GeV (right).
\label{figmass}}
\end{figure}

The Higgs Mass $M_{\mathrm H}$ is the only unknown parameter left in the SM.
Once its mass is known, the profile of the Higgs sector in the 
SM is uniquely determined. High precision is needed in order to discriminate between
the SM and theories with an extended Higgs sector {\it e.g.} SUSY models. 
At the LC the Higgs mass can be 
best measured exploiting the kinematics in the Higgs-strahlung process 
${\mathrm e^+e^- \rightarrow Z \rightarrow Z H}$. The use of the recoil mass spectrum for leptonic 
Z decays ${\mathrm (Z \rightarrow e^+e^-, \mu^+\mu^-)}$ yields an accuracy of 110 MeV for 
500\,fb$^{-1}$, without any requirement on the Higgs boson decay~\cite{hmass}. The accuracy can be 
improved by identifying and reconstructing the Higgs decay mode and applying
kinematic fits with constraints. For $M_{\mathrm H}\le$ 130\,GeV the dominant decay mode is into
${\mathrm b\bar{b}}$, whereas for higher masses the ${\mathrm WW^*}$ channel becomes more important.
Examples of reconstructed mass spectra are shown in Fig.~\ref{figmass}.
The combination of these measurements yields an accuracy between 40 and 80\,MeV for
$M_{\mathrm H}$ between 120 and 180\,GeV for ${\cal L}$ = 500\,fb$^{-1}$~\cite{hmass}.
%%%%%%%%%%%%%%%%%%%%%%%%%%%%%%%%%%%%%%%%%%%%%%%%%%%%%%%%%%%%%%%%%%%%%%%%%%%%%%%%%%%%%%%%%%%%%%
\subsection{Quantum Numbers}
\begin{figure}[htb]
\center{
\epsfig{figure=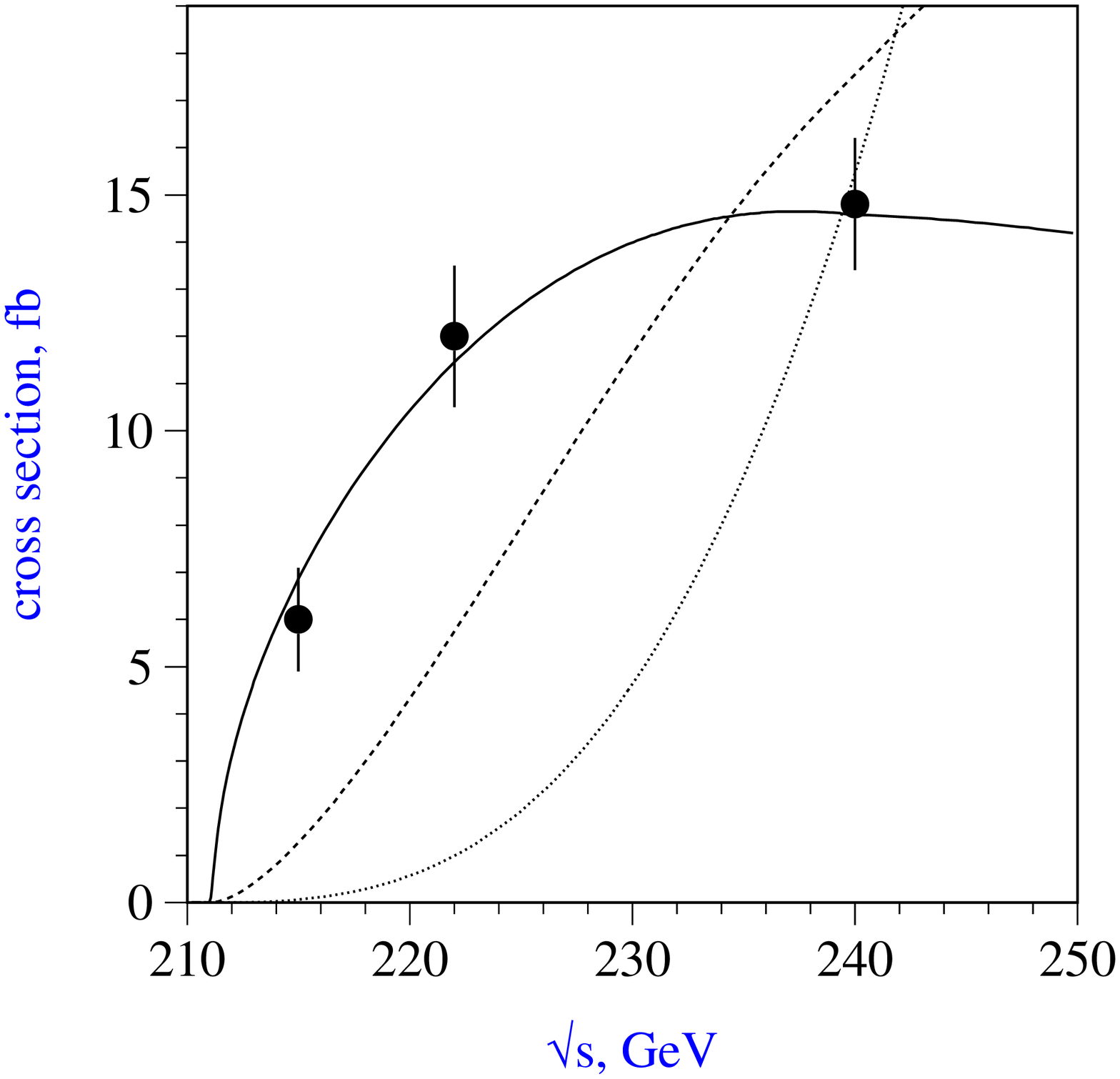,height=5.5cm,width=5.5cm}
\hspace{2cm}
\epsfig{figure=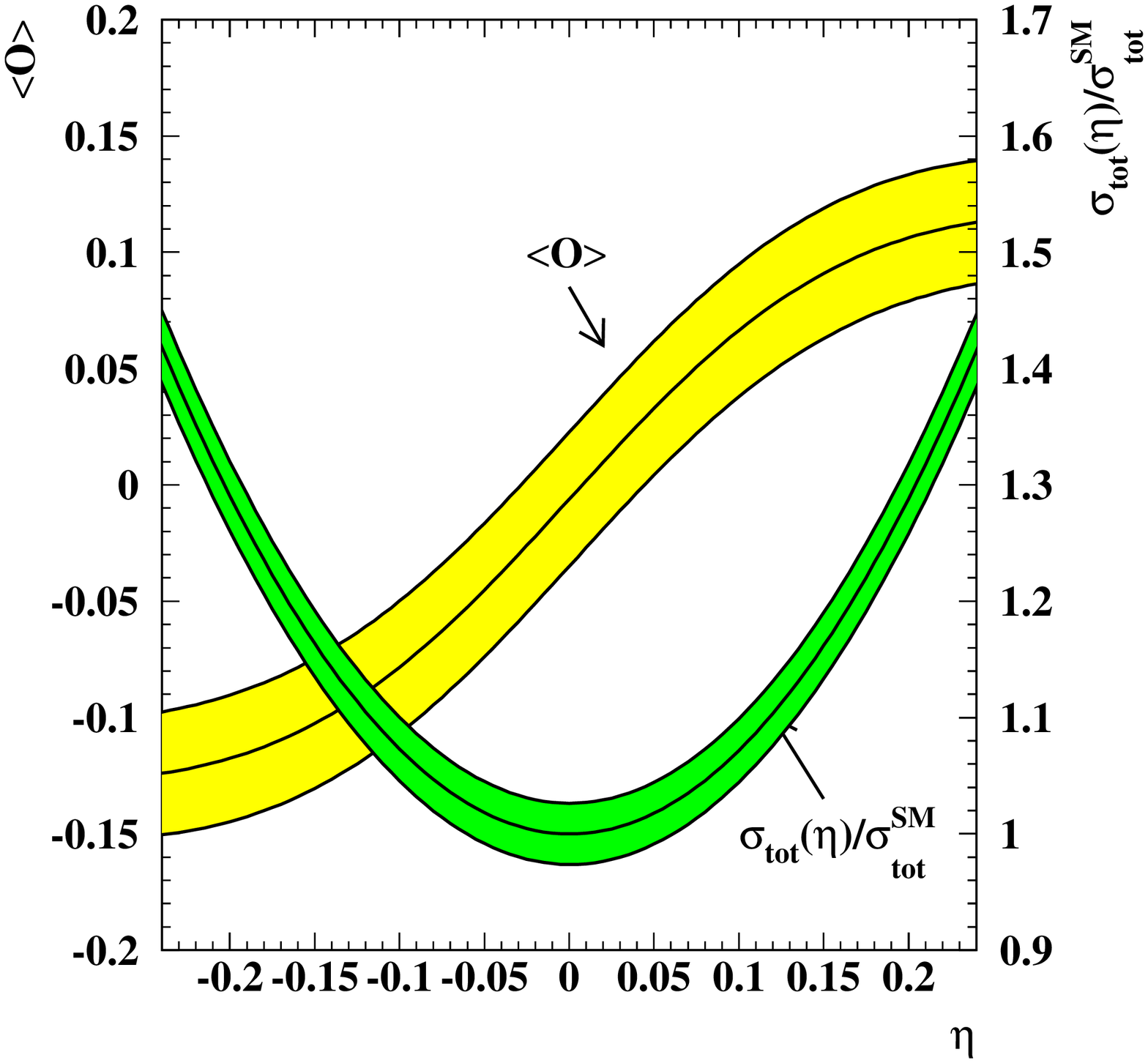,height=5.5cm,width=5.5cm}}
\caption{
Left: Simulated measurement of the ${\mathrm e^+e^- \rightarrow ZH}$ cross-section 
for $M_{\mathrm H} = 120$\,GeV with 20\,fb$^{-1}$ at three center-of-mass energies compared
to predictions for a spin-0 (full line), and examples of spin-1 (dashed line)
and spin-2 (dotted line) particles. Right: The dependence of the expectation value of the optimal observable (light grey) and the total cross-section (dark grey) on $\eta$ for 
$M_{\mathrm H}$ = 120\,GeV, $\sqrt{s}$ = 350\,GeV and ${\cal L}$  = 500\,fb$^{-1}$. The shaded bands
show the the 1$\sigma$ uncertainties expected. 
\label{figqn}}
\end{figure}

The spin, parity and charge-conjugation quantum numbers of the Higgs bosons
can be determined at a LC in a model-independent way. The observation of Higgs boson production at the photon collider ${\mathrm \gamma\gamma\rightarrow H}$ or of the decay 
${\mathrm H \rightarrow \gamma \gamma}$ would rule 
out $J=1$ and require $C$ to be positive. An scan of the threshold rise of the 
Higgs-strahlung cross-section and the measurement of the  angular dependence of
the Higgs production allow to determine $J$ and $P$ uniquely and to distinguish between
a $CP$-even SM like Higgs boson ($0^{++}$), a $CP$-odd ($0^{+-}$) state A, and a $CP$-violating mixture denoted by $\Phi$ of the two~\cite{miller}.
A threshold scan with a luminosity of 20\,fb$^{-1}$ at three center-of-mass 
energies is sufficient to distinguish between the different rise 
expected for different spin assumptions~\cite{spin}~(see Fig.~\ref{figqn} (left)).
In a general model with two Higgs doublets (2HDM) the three neutral Higgs bosons 
correspond to arbitrary mixtures of $CP$ eigenstates and their production and decay may 
exhibit $CP$ violation. In this case, the amplitude for the Higgs-strahlung process can 
be described by adding a ZZA coupling with strength $\eta$ to the SM matrix element.
The squared amplitude is then given by
$|{\cal M}_{\mathrm Z\Phi}|^2 = |{\cal M}_{\mathrm ZH}|^2 + 2 \eta 
Re({\cal M}_{\mathrm ZH}^*{\cal M}_{\mathrm ZA}) +
\eta^2 |{\cal M}_{\mathrm ZA}|^2$~\cite{DK}. The first term corresponds to the SM 
cross-section, the second, linear in $\eta$, to the interference term, gives rise 
to $CP$-violating effects 
% {\it e.g.} a forward-backward asymmetry in the Higgs production, 
and the third term,
quadratic in $\eta$ and $CP$ conserving, increases the total cross-section. 
The latter two change also the angular distributions of the decay 
${\mathrm Z \rightarrow f\bar{f}}$. The information carried by these distributions has 
been 
analysed using the formalism of optimal observables for the 
${\mathrm ZH \rightarrow \mu^+\mu^-X}$ final state for $M_{\mathrm H}$ = 120\,GeV, 
$\sqrt{s}$ = 350\,GeV and ${\cal L}$ = 500\,fb$^{-1}$. The accuracy on $\eta$ has been 
found to be  $\approx$ 0.03~\cite{cp}.
%%%%%%%%%%%%%%%%%%%%%%%%%%%%%%%%%%%%%%%%%%%%%%%%%%%%%%%%%%%%%%%%%%%%%%%%%%%%%%%%%%%%%%%%%%%%%%
\subsection{Couplings to Electroweak Gauge Bosons}
The couplings to massive gauge bosons W and Z can be probed independently 
and best in the measurement of the production cross-sections for Higgs-strahlung
(${\mathrm e^+e^- \rightarrow Z \rightarrow ZH}$), 
which is proportional to $g^2_{\mathrm HZZ}$, and
WW--fusion (${\mathrm e^+e^- \rightarrow H^0 \nu_e \bar{\nu}_e}$), which is proportional 
to $g^2_{\mathrm HWW}$.

\begin{figure}[htb]
\center{
\epsfig{figure=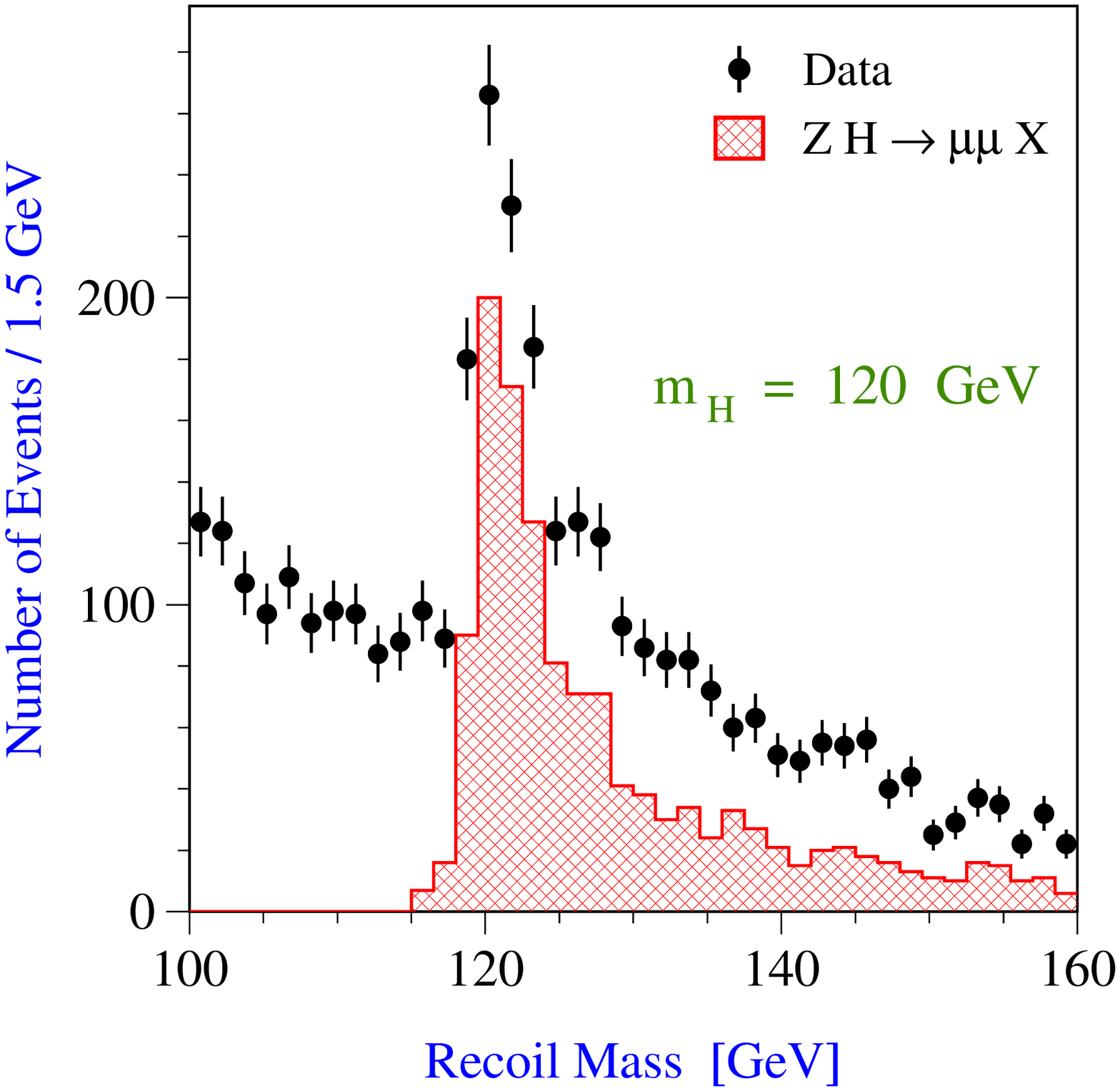,height=5.5cm,width=5.5cm}
\hspace{2cm}
\epsfig{figure=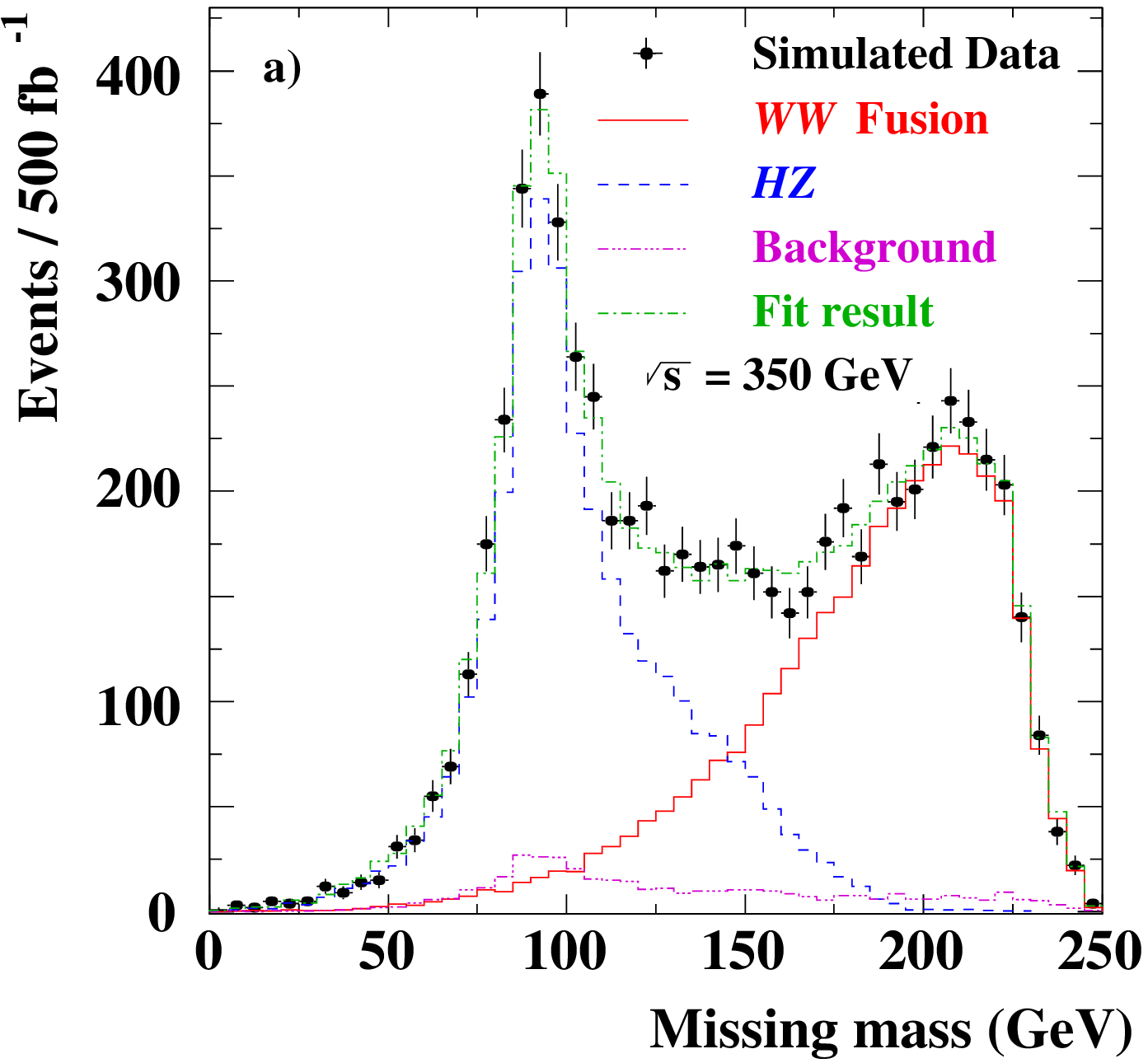,height=5.5cm,width=5.5cm}}
\caption{Left: The $\mu^+\mu^-$ recoil mass distribution in the process 
${\mathrm e^+e^- \rightarrow ZH \rightarrow X \mu^+\mu^-}$ for 
$M_{\mathrm H}$ = 120 \,GeV and 500\,fb$^{-1}$ at $\sqrt{s}$ = 350\,GeV.
Right: Simulation of the missing mass distribution in ${\mathrm b\bar{b}\nu\bar{\nu}}$ events
for 500\, fb$^{-1}$ at $\sqrt{s}$ = 350\,GeV.
\label{figcross}}
\end{figure}

The cross-section for the Higgs-strahlung process can be measured by analysing the 
mass spectrum of the system recoiling against leptonic decays of the Z 
(see Fig.~\ref{figcross} (left)).
This provides a cross-section determination independent of the Higgs boson decay modes.
The accuracy achieved in detailed studies is between 2.5\% and 3.0\% for Higgs masses 
between 120 and 160\,GeV~\cite{hmass}. 

The cross-section for WW--fusion can be determined in the ${\mathrm b\bar{b}\nu\bar{\nu}}$
final state, where these events can be well separated from the Higgs-strahlung final
state  and the background processes by exploiting the different spectra of the 
${\mathrm \nu\bar{\nu}}$ invariant mass (see Fig.~\ref{figcross} (right)). From a 
simultaneous fit of the above contributions the cross-section can be extracted with an 
accuracy between 3\% and 13\% for $M_{\mathrm H}$ between 120 and 160\,GeV~\cite{hww}. 

The measurement of the branching ratio ${\mathrm H\rightarrow WW^*}$ 
provides an alternative 
means to access the $g_{\mathrm HWW}$ coupling. 
The experimentally study has been performed for ${\mathrm Z\rightarrow \ell^+\ell^-}$, ${\mathrm H\rightarrow q\bar{q}^\prime q\bar{q}^\prime}$ and ${\mathrm Z\rightarrow q\bar{q}}$, ${\mathrm H\rightarrow q\bar{q}^\prime l\nu}$  final states achieving a  precision of 5 to 2\% for Higgs masses between 120 and 160\,GeV~\cite{brww}.

The loop-mediated coupling to photons can be measured 
best at the photon collider in the reaction ${\mathrm \gamma\gamma \rightarrow H}$ 
with an accuracy
of 2\% for 120\,GeV SM like Higgs boson with 150\,fb$^{-1}$~\cite{hgamma1}. 
An alternative method is given by the measurement of 
${\mathrm BR}({\mathrm H\rightarrow \gamma \gamma})$. Due to the its small value
in the SM of  2$\times$10$^{-3}$ the precision for $M_{\mathrm H}$ = 120\,GeV 
is 23\%(16\%) for 
500 (1000)\,fb$^{-1}$ exploiting the $\nu\bar{\nu}\gamma\gamma$ and ${\mathrm q\bar{q}\gamma\gamma}$ final states~\cite{hgamma2} at $\sqrt{s}$ = 500\,GeV.

%%%%%%%%%%%%%%%%%%%%%%%%%%%%%%%%%%%%%%%%%%%%%%%%%%%%%%%%%%%%%%%%%%%%%%%%%%%%%%%%%%%%%%%%%%%%%%
\subsection{Couplings to Fermions and Gluons}
The accurate determination of the Higgs couplings to fermions is important as a proof of the 
Higgs mechanism to be responsible for generating the fermion masses. The partial decay widths to 
fermions are poportional to $g_{{\mathrm Hf\bar{f}}}^2 = m_{\mathrm f}^2/v^2$ 
and are fully determined by the fermions mass in the SM. An observation of deviations from the SM expectations probes
the parameters of an extended Higgs sector. The accuracy of the branching ratio measurements 
relies on the excellent performance of the envisaged vertex detector.
In the hadronic Higgs decay channels the fractions of ${\mathrm b\bar{b}}$, 
${\mathrm c\bar{c}}$ and ${\mathrm gg}$ final states  are extracted by a binned 
maximum likelihood fit to the jet flavour 
tagging probabilities for the Higgs boson decay candidates, while the $\tau^+\tau^-$ final 
states are selected by a dedicated likelihood, based mainly on vertexing and calorimetric 
response. These measurements are sensitive to the product 
$\sigma_{\mathrm ZH,\nu\bar{\nu}H} \times {\mathrm BR}({\mathrm H \rightarrow f\bar{f}})$. 
Using the results discussed
above for the production cross-sections the branching ratios can be determined to
the following relative accuracies for $M_{\mathrm H}$ = 120 (140)\,GeV for 500\,fb$^{-1}$:
${\mathrm b\bar{b}}$ 2.4 (2.6)\%,  ${\mathrm c\bar{c}}$ 8.3 (19.0)\%, ${\mathrm gg}$ 5.5 (14.0)\%, $\tau^+\tau^-$ 5.0 (8.0)\%~\cite{br}
(see also Fig.~\ref{figbr} (left)).

The Higgs coupling to the top quark is the largest coupling in 
the SM ($g_{\mathrm Ht\bar{t}}^2 \simeq 0.5$ to be compared with 
$g_{\mathrm Hb\bar{b}}^2 
\simeq 4 \times 10^{-4}$). However, for a light Higgs boson this coupling is 
accessible indirectly in the loop process $\mathrm{H \rightarrow g g}$ and directly 
only in the Yukawa process ${\mathrm e^+e^- \rightarrow t \bar t H}$~\cite{tth1}. 
This process has a cross-section of the order of only 2.5\,fb at $\sqrt{s}$ = 800\,GeV,
including QCD corrections~\cite{tth1b}.
%The distinctive signature, consisting  of two W bosons and four b-quark jets, 
%makes it possible to isolate these events from the thousand times larger backgrounds. 
For an integrated luminosity of 1000\,fb$^{-1}$ the uncertainty in the Higgs top Yukawa coupling 
is 5.5\%~\cite{tth2}. 
For $M_{\mathrm H} \ge 2 \times m_{\mathrm t}$ the Higgs top Yukawa couplings can be measured from the 
${\mathrm H \rightarrow t \bar t}$ branching fractions. A study has been performed based
on the analysis of the process ${\mathrm e^+e^- \rightarrow \nu_e \bar \nu_e H 
\rightarrow \nu_e \bar \nu_e t \bar t}$ for 350\,GeV$<M_{\mathrm H}<$ 500\,GeV
at $\sqrt{s}$ = 800\,GeV yielding an accuracy of 5\% (12\%) for $M_{\mathrm H}$ = 400 (500)\,GeV for an integrated
luminosity of 500\,fb$^{-1}$~\cite{tth3}. 
%%%%%%%%%%%%%%%%%%%%%%%%%%%%%%%%%%%%%%%%%%%%%%%%%%%%%%%%%%%%%%%%%%%%%%%%%%%%%%%%%%%%%%%%%%%%%%
\subsection{Extraction of Higgs Couplings}
\begin{figure}[htb]
\center{
\epsfig{figure=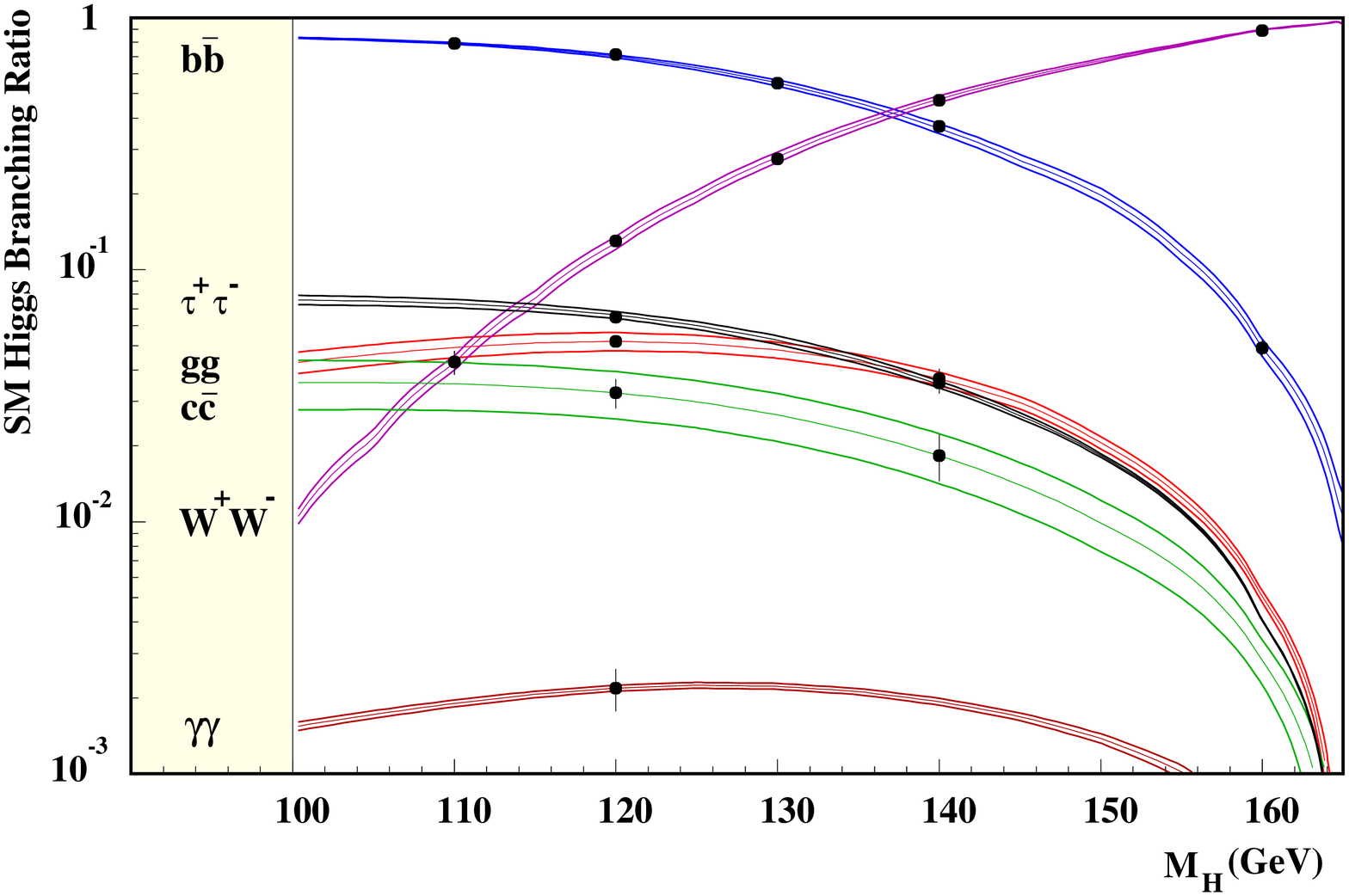,height=5.5cm,width=5.5cm}
\hspace{2cm}
\epsfig{figure=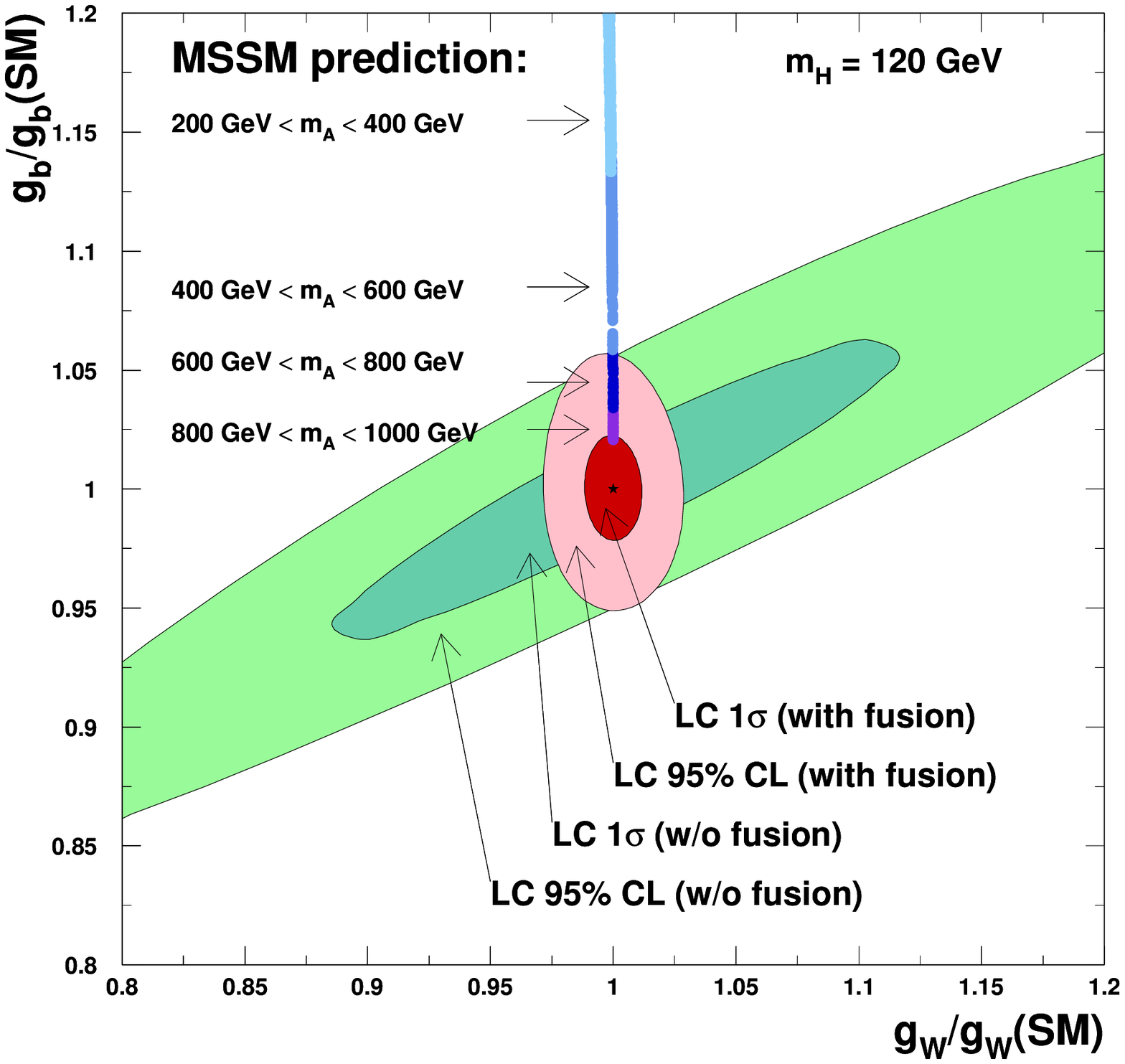,height=5.5cm,width=5.5cm}}
\caption{Left: The predicted SM Higgs boson branching ratios. Point with error bars show the 
expected experimental accuracy, while the lines shows the estimated uncertainties on the SM 
predictions.  Right: Higgs coupling determination from a global fit compared to deviations expected in the MSSM.
\label{figbr}}
\end{figure}

The Higgs boson production and decay rates discussed above, can be used to 
measure the Higgs couplings to gauge bosons and fermions. 
After the Higgs boson is discovered, this is the first crucial step in
establishing experimentally the Higgs mechanism for mass generation.
Since some of the couplings
of interest can be determined independently by different observables while 
other determinations are partially correlated, it is interesting to perform
a global fit to the measurable observables and to extract the Higgs couplings
in a model--independent way.
This method optimises the available information and can take properly
into account the experimental correlation between different measurements.
%A dedicated program, {\sc HFitter}~\cite{hfitter} has been developed based on 
%the {\sc Hdecay}~\cite{hdecay} program for the calculation of the Higgs boson
%branching ratios. The following inputs have been used: $\sigma_{HZ}$, 
%$\sigma_{H\nu\bar{\nu}}$, {\mathrm BR}($H\rightarrow WW^*$), BR($H \rightarrow \gamma 
%\gamma$), BR($H \rightarrow b \bar b$), BR($H \rightarrow \tau^+ \tau^-$),
%BR($H \rightarrow c \bar c$), BR($H \rightarrow g g$), $\sigma_{t \bar t H}$.
%%For correlated measurements the full covariance matrix has been used.
The relative uncertainties on the Higgs boson couplings from this global fit
as calculated by {\sc HFitter}~\cite{hfitter} are given for $M_{\mathrm H}$ = 120\,GeV\  and 140\,GeV\  and 
500\,fb$^{-1}$ in Table~\ref{tabhfitter} (see also Fig.~\ref{figbr} (right)).
\begin{table}[htb!]
\label{tabhfitter}
\caption{Relative accuracy on Higgs boson couplings obtained from a global fit.}
\vspace{0.15cm}
\begin{center}
\begin{tabular}{|l|c|c|c|c|c|c|}
\hline
Coupling & $g_{\mathrm HWW}$ & $g_{\mathrm HZZ}$ & $g_{\mathrm Ht\bar{t}}$ & $g_{\mathrm Hb\bar{b}}$ & $g_{\mathrm Hc\bar{c}}$ & $g_{\mathrm H\tau\tau}$\\
\hline
$M_{\mathrm H}$  = 120\,GeV & 0.012 & 0.012 & 0.030 & 0.022 & 0.037 &  0.033 \\
$M_{\mathrm H}$  = 140\,GeV & 0.020 & 0.013 & 0.061 & 0.022 & 0.102 &  0.048 \\ \hline
\end{tabular}
\end{center}
\end{table}
%%%%%%%%%%%%%%%%%%%%%%%%%%%%%%%%%%%%%%%%%%%%%%%%%%%%%%%%%%%%%%%%%%%%%%%%%%%%%%%%%%%%%%%
\subsection{The Total Higgs Width}
The total decay width of the Higgs boson is predicted to be too narrow
to be resolved experimentally for Higgs boson masses below the ZZ--threshold.
Above approximately 200\,GeV the total width can be measured directly.
For lower masses, indirect methods, exploiting relations between the total
decay width and the partial widths for exclusive final states, can be applied:
$\Gamma_{\mathrm tot}=\Gamma_{\mathrm X}/\mathrm{BR}({\mathrm H \rightarrow  X})$.
Two feasible options exist for light 
Higgs bosons: i) the extraction of $\Gamma_{\mathrm WW}$ from the measurement of the 
${\mathrm WW}$--fusion cross-section combined with the measurement of 
$\mathrm{BR}({\mathrm H\to WW^*})$ and ii) the measurement of the ${\mathrm \gamma\gamma\to H}$ 
cross-section at a $\gamma\gamma$ collider combined with the measurement of 
$\mathrm{BR}({\mathrm H\to \gamma\gamma})$ in ${\mathrm e^+e^-}$ collisions.
The ${mathrm WW}$--fusion option yields a precision on $\Gamma_{\mathrm tot}$ of 6\% to 13\% for
Higgs boson masses between 120 and 160\,GeV, while the $\gamma\gamma$ option
yields a larger error (23\% for $M_{\mathrm H}$ = 120\,GeV) dominated by the large uncertainty in the 
$\mathrm{BR}({\mathrm H\to \gamma\gamma})$ determination discussed above.
Assuming the $SU(2)\times U(1)$ relation $g_{\mathrm HWW}^2/g_{\mathrm HZZ}^2 =
1/\cos^2{\theta_W}$ to be valid, the measurement of the Higgs--strahlung
cross-section provides a viable alternative with potentially higher
mass reach than the ${\mathrm WW}$--fusion option. The errors on $\Gamma_{\mathrm tot}$ can then be reduced to below 4\% for $M_{\mathrm H}$ between 140 and 160\,GeV.
%%%%%%%%%%%%%%%%%%%%%%%%%%%%%%%%%%%%%%%%%%%%%%%%%%%%%%%%%%%%%%%%%%%%%%%%%%%%%%%%%%%%%%%%%%%%%
\subsection{The Higgs Potential}
In order to fully establish the Higgs mechanism, the Higgs potential 
$V = \lambda~(|\phi|^2-\frac{1}{2}v^2)^2$ with $v = (\sqrt{2}G_{\mathrm F})^{-1/2} 
\simeq$ 246\,GeV must be reconstructed through the determination of the 
triple, $\lambda_{\mathrm HHH}$, and quartic, $\lambda_{\mathrm HHHH}$, Higgs self couplings. 
While effects from the quartic coupling may be too small to be observed at the
LC, the triple Higgs coupling can be measured in the double Higgs boson 
production processes ${\mathrm e^+e^- \rightarrow HHZ}$ and ${\mathrm \nu_e\bar\nu_eHH}$.
 In 
${\mathrm e^+e^-}$ collisions up to 1~TeV the double Higgs boson associated production 
with the $Z$ is favoured~\cite{hpot1}. The sensitivity to $\lambda_{\mathrm HHH}$ from the measurement
of $\sigma_{\mathrm HHZ}$ and $\sigma_{\mathrm \nu\bar\nu HH}$ is diluted by the effects of 
other diagrams, not involving the triple Higgs coupling but leading
to the same final state. The huge backgrounds and the small
signal cross-section ($\sigma$ = 0.18 fb for $M_{\mathrm H}$ = 120\,GeV 
and $\sqrt{s}$ = 500\,GeV) make this measurement an experimental challenge. 
A determination of $\lambda_{\mathrm HHH}$ with a statistical 
accuracy of 22\% for $M_{\mathrm H}$ = 120\,GeV is possible with 1000\,fb$^{-1}$~\cite{hpot2}. 
In the SM the value of $\lambda_{\mathrm HHH}$ is fixed  after the measurement of the Higgs
mass. However, in models with an extra Higgs doublet, additional trilinear Higgs 
couplings are also present such as $\lambda_{\mathrm hhH}$, $\lambda_{\mathrm hhA}$, 
$\lambda_{\mathrm hhh}$ and $\lambda_{\mathrm HAA}$, which depend also on 
$\tan \beta$ and $M_{\mathrm A}$  and will change the shape of the potential~\cite{hpot3}.
%%%%%%%%%%%%%%%%%%%%%%%%%%%%%%%%%%%%%%%%%%%%%%%%%%%%%%%%%%%%%%%%%%%%%%%%%%%%%%%%%%%%%%%%%%%%%
\section{SUSY Higgs Bosons}
Several extension of the SM model introduce additional Higgs doublets and singlets.
A no loose theorem \cite{gunion} guarantees that in a general SUSY model embedded
in a GUT scenario at least one Higgs boson will be observable at $\sqrt{s}$ = 500\,GeV with 
${\cal L}$ = 500\,fb$^{-1}$. A specific scenario studied in some detail is the MSSM.
The mass reach for pair production  of ${\mathrm HA}$ and ${\mathrm H^+H^-}$ extends to $\approx \sqrt{s}/2 -\epsilon$. 
This mass range is extended
for single production of $\mathrm H$ and $\mathrm A$ at a photon collider up to 0.8 $\times \sqrt{s_{\mathrm ee}}$.
Establishing the existence of these additional Higgs bosons and the determination of 
their masses and main decay modes represents an important part of the LC physics program.
%%%%%%%%%%%%%%%%%%%%%%%%%%%%%%%%%%%%%%%%%%%%%%%%%%%%%%%%%%%%%%%%%%%%%%%%%%%%%%%%%%%%%%%%%%%%%
\subsection{Direct Determination of SUSY Higgs Boson Properties}
The decay channels ${\mathrm H,A \rightarrow b \bar{b}, H^+ \rightarrow t \bar{b}}$ or 
${\mathrm W^+ h}$, 
${\mathrm h \rightarrow  b\bar{b}}$ will provide with very distinctive 4~jet and 8~jet final states with 
4~$\mathrm b$-quark jets that can be efficiently identified and reconstructed. Exemplificative 
analyses have been performed for these channels, showing that an accuracy of about 0.3\% 
on their mass and of $\simeq 10\%$ on $\sigma \times BR$ can be obtained~\cite{hpm}. 
In SUSY models additional decay channels may open. Particularly interesting is the
decay of neutral Higgs bosons into an invisible final state {\it e.g.} 
${\mathrm h \rightarrow \chi^0\chi^0}$. The comparison of the number of events observed 
in the ${\mathrm e^+e^-\rightarrow ZH \rightarrow \ell^+\ell^- X}$ final state with the sum over the 
branching ratios of the visible decay modes allows an indirect determination of 
${\mathrm BR(H \rightarrow inv.)}$ with an accuracy of better than 20\% for 
${\mathrm BR(H \rightarrow inv.) \ge}$ 0.05.
%%%%%%%%%%%%%%%%%%%%%%%%%%%%%%%%%%%%%%%%%%%%%%%%%%%%%%%%%%%%%%%%%%%%%%%%%%%%%%%%%%%%%%%%%%%%%
\subsection{Indirect Determination of SM/MSSM Nature of a Light Higgs Boson}
\begin{figure}[htb]
\center{
\epsfig{figure=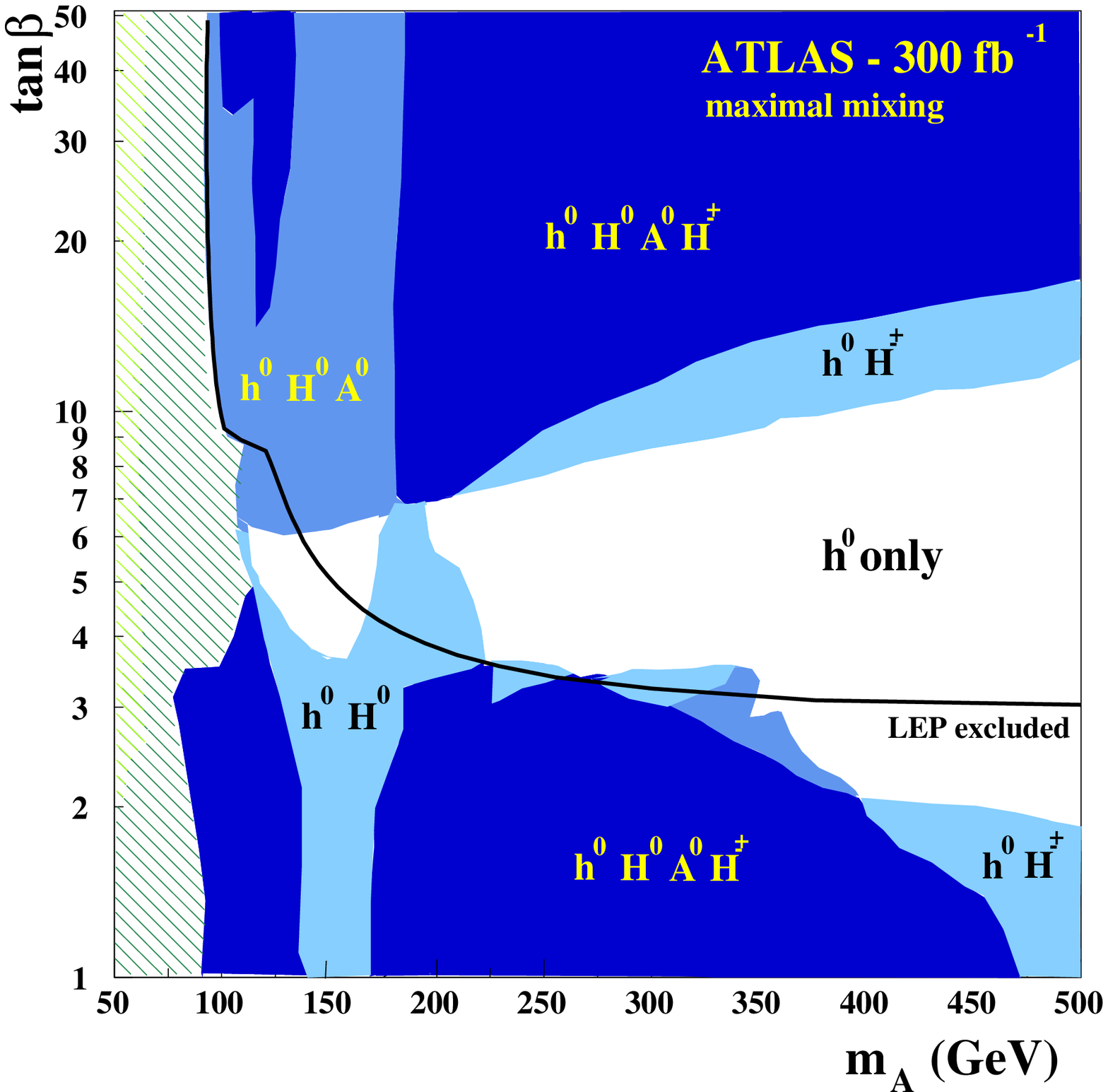,height=5.5cm,width=5.5cm}
\hspace{2cm}
\epsfig{figure=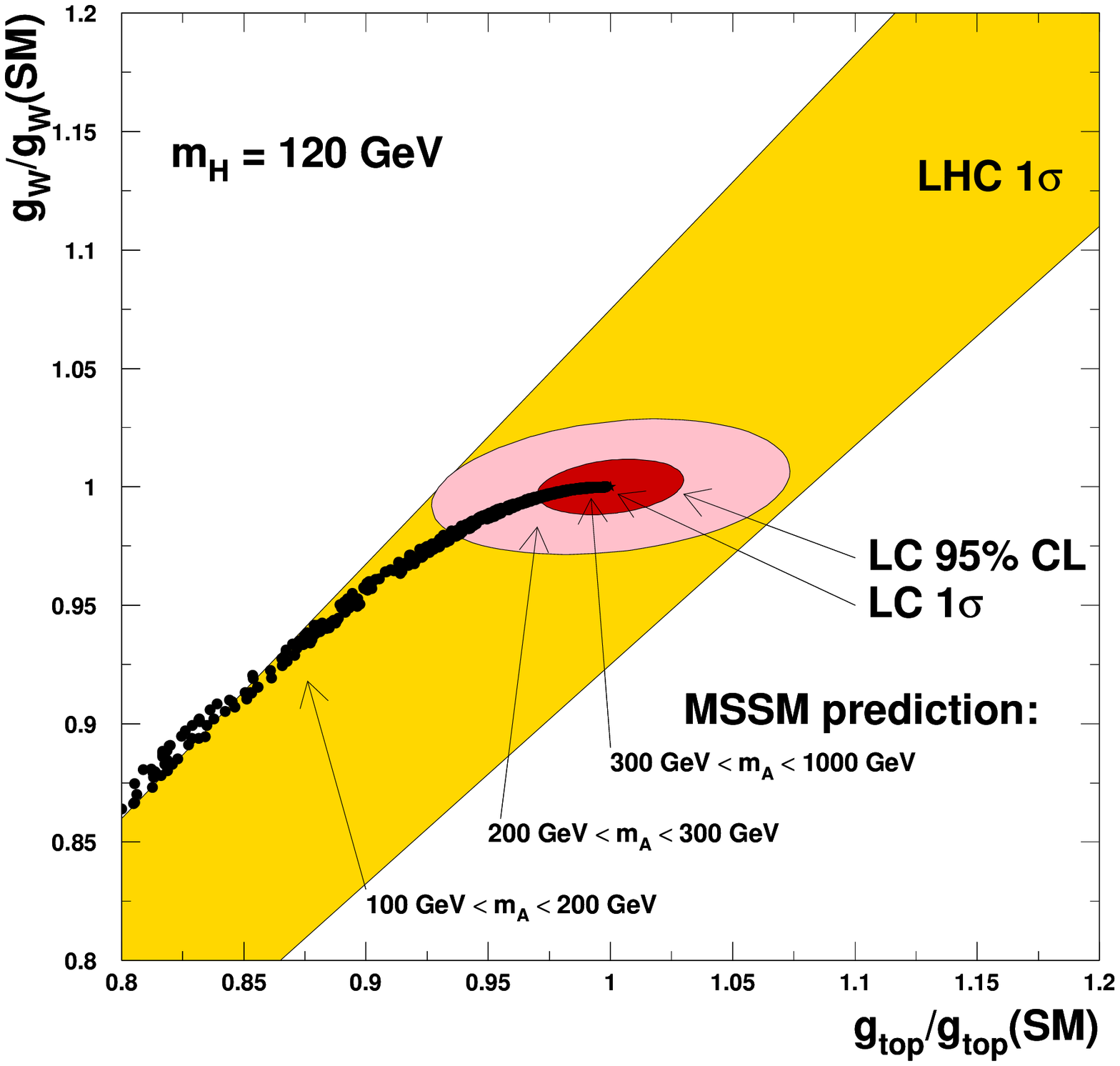,height=5.5cm,width=5.5cm}}
\caption{
Left: Higgs bosons which are observable in ATLAS with 300\,fb$^{-1}$ 
in the maximal mixing scenario of the MSSM. 
In the white region only the lightest boson 
is obervable if only SM-like decays are accessible. At the LC the SM or MSSM nature of the ${\mathrm h}$ boson can be distinguished from the precision measurements for the entire area.
Right: A comparison of the accuracy in the determination of the $g_{\mathrm ht\bar{t}}$
and $g_{\mathrm HWW}$ Higgs couplings at the LHC and the LC compared to the predictions from the MSSM.
\label{figlhc}}
\end{figure}

The discovery of a neutral Higgs boson, with mass in the range 115\,GeV $< M_{\mathrm H} 
<$ 140\,GeV, will raise the question of whether the observed particle is the SM
Higgs boson or the lightest boson from the Higgs sector of a SM extension.
It has been shown that, for a large fraction of the $\tan \beta - M_{\mathrm A}$ 
parameter plane in the MSSM, this neutral boson will be the only Higgs state
observed at the LHC (see Fig.~\ref{figlhc} (left)). In this circumstance, a Higgs particle 
generated by a complex multi-doublet model could be indirectly recognised only by a 
study of its couplings. If the ${\mathrm HZZ}$ coupling, measured by the Higgs-strahlung production 
cross-section independently from the Higgs boson decay mode, turns out to be 
significantly smaller than the SM expectation, this will signal the existence 
of extra Higgs doublets.

The determination of the Higgs boson decay branching ratios with the accuracy
anticipated by these studies can be employed to identify the SM or MSSM 
nature of a light neutral Higgs boson, because the Higgs boson decay widths
to a specific final state are modified by factors involving 
$\tan\beta$ and  $\alpha$ in the MSSM compared to the SM.
Therefore, deviations in the ratios of branching ratios such as
{\it e.g.} $\frac{\mathrm BR(h \rightarrow W W^*)}{\mathrm BR(h \rightarrow b \bar b)}$
from their SM expectations can reveal the MSSM nature of the Higgs boson and also provide 
indirect information on the mass of the $CP$-odd $A^0$ Higgs boson, even when
it is so heavy that it can not be directly observed at $\sqrt{s}$ = 500\,GeV. 

To compare the SM predictions with those in MSSM, a complete scan of the MSSM 
parameter space has been performed for $M_{\mathrm h} = (120 \pm 2)$\,GeV.
It was found that for $M_{\mathrm A} \le$ 600\,GeV
95\% of the MSSM solutions can be excluded at 95\% confidence level~\cite{br,carena}.
In the case of an significant deviation from the SM expectation this can 
be translated into an indirect determination of $M_{\mathrm A}$ yielding an accuracy of 70 to 100\,GeV for 300\,GeV $< M_{\mathrm A} <$ 600\,GeV.
%%%%%%%%%%%%%%%%%%%%%%%%%%%%%%%%%%%%%%%%%%%%%%%%%%%%%%%%%%%%%%%%%%%%%%%%%%%%%%%%%%%%%%%%%%%%%%%%%%%%
\section{The Complementarity with the LHC}
At the LHC the SM Higgs boson, or at least one Higgs boson in the MSSM,
will be observed. Beyond its discovery a limited number of measurements
of Higgs boson properties can be carried out at the LHC (mass, total width for 
a heavy Higgs boson, some ratios of couplings). Further perspectives for the observation 
of an invisible decaying Higgs,  and measurements of the total width for lower masses
and of the $g_{\mathrm HWW}$ coupling have been recently suggested for the LHC~\cite{zeppenfeld}. 

The complementarity of the linear collider data to the picture of the Higgs 
sector as it will have been outlined by the LHC is therefore threefold.
First the accuracy of those measurements, which are possible at the LHC, can be 
significantly increased for e.g. $M_{\mathrm H}$ =120\ (160)\,GeV:
$\Delta\,M/M$ = 9 (10)$\times$10$^{-4}$ at the LHC and  3 (4)$\times$10$^{-4}$ at the LC,
$g_{\mathrm Ht\bar{t}}/g_{\mathrm HWW}$ = 0.070 at the LHC and 0.023 at the LC, 
$g_{\mathrm Ht\bar{t}}/g_{\mathrm HWW}$ = 0.050 at the LHC and 0.022 at the LC.
Secondly the absolute measurements of all the relevant 
Higgs boson couplings, including the Higgs self coupling, will be possible 
only at the LC. Finally extended Higgs sector scenarios (e.g. 
invisible Higgs boson decays or 2HDM) can be observed at the linear collider 
closing the loopholes of a possible non-discovery at the LHC.

\section{Summary}
The search for the Higgs boson and the study of its properties is one of the 
main goals of present research in particle physics. The central role of a
linear collider in the understanding of the mechanism of electroweak symmetry 
breaking, complementing the data to be acquired at the Tevatron and at the LHC, has been clearly outlined by the studies 
carried out world-wide.
\section*{Acknowledgements}
I am grateful to the organisers especially to J.\,Tran\,Thanh\,Van for their
invitation and the inspiring atmosphere. The results reported here are due to 
the activities of many colleagues, in the regional Higgs working groups of the 
Worldwide Study on Physics at Linear Colliders, whose contributions are
gratefully acknowledged. 

\end{document}